\documentclass[a4paper]{jpconf}
\usepackage{graphicx}
\begin{document}
\title{Searching for continuous gravitational wave signals using LIGO and Virgo detectors}  

\author{Paola Leaci (for the LIGO Scientific Collaboration and the Virgo Collaboration)}

\address{Max-Planck-Institut f\"{u}r Gravitationsphysik, Albert-Einstein-Institut, 14476 Golm, Germany}

\ead{paola.leaci@aei.mpg.de}

\begin{abstract}
Direct and unequivocal detection of gravitational waves represents a great challenge of contemporary physics and astrophysics. A worldwide effort is currently operating towards this direction, building ever sensitive detectors, improving the modelling of gravitational wave sources and employing ever more sophisticated and powerful data analysis techniques. 
In this paper we review the current status of LIGO and Virgo ground based interferometric detectors and some data analysis tools used in the continuous wave searches to extract the faint gravitational signals from the interferometric noise data. Moreover we discuss also relevant results from recent continuous wave searches.
\end{abstract}

(Some figures in this article are in colour only in the electronic version)

\section{Introduction}

Gravitational waves, predicted to exist by Einstein's General Theory of Relativity~\cite{EinstGW}, are ripples in space-­time propagating at light speed and produced by non-axially symmetric mass accelerations, by analogy with electric charges in any accelerated motion that emit electromagnetic waves (travelling at light speed). However, gravitational waves are quite different from electromagnetic waves. They are both transverse waves, but gravitational waves are characterized by two polarization states, denoted as ``+'' and ``$\times$'', that differ by a rotation of $45$ degrees around the propagation axis, demonstrating the quadrupolar (spin-2) nature of the gravitational radiation. On the contrary, the two polarization states of electromagnetic waves differ by a rotation of $90$ degrees, reflecting thus the dipolar (spin-1) characteristics of the electromagnetic radiation. The emission mechanisms are also quite different: gravitational waves result from the coherent emission from bulk motions of energy, while electromagnetic waves result from an incoherent superposition of waves from molecules, atoms and particles. The amount of energy radiated as gravitational waves by any mechanical system constructed by man is so small that it will probably never be observed. For this reason we hope to observe gravitational radiation emitted by sources at astrophysical distances. 
Indeed, even though gravitational waves have not yet been directly detected by any detectors, a very strong indirect proof of their existence was given by the observation of the binary pulsar PSR B1913+16, discovered in 1974 by the astronomers R. Hulse and J. Taylor~\cite{HulseTaylor}. Using the giant radio telescope at the Arecibo Observatory of Puerto Rico~\cite{Arecibo} to search systematically for pulsars, they observed that a particular pulsar (PSR B1913+16) was changing its motion rapidly and that the variation in pulse rate was caused by the changing Doppler effect. The decrease of the orbital period of such a pulsar around its companion could only be explained if angular momentum and energy were carried away from this system by gravitational waves.

Because of the extreme weakness of the interaction of gravitational radiation with matter, gravitational waves travel almost undisturbed  from astrophysical sources to Earth, without being scattered or absorbed by interstellar dust and debris, carrying thus astronomical information which electromagnetic waves do not carry.
An analysis of gravitational radiation would provide information of great value about the inaccessible and remote locations of the cosmos. It would tell us something about the behaviour of space-time and matter under the most extreme conditions, and it would also provide a check on Einstein's General Theory of Relativity. The detection of gravitational waves will help us to understand the dynamics of large-scale events in the Universe, like the death of whole stars, the explosion of quasars, the birth and the collisions of Black Holes (BHs for short; for more details see for instance Ref.~\cite{LeaciPhD} and references therein).

Nowadays laser interferometry is the basis of the most sensitive gravitational wave detectors, such as the LIGO (Laser Interferometer Gravitational wave Observatory) and Virgo detectors~\cite{LIGOref1,LIGOref2,Virgo}, whose current performances are concisely discussed in the next section.

The remainder of the paper is organized as follows. In Sec.~\ref{GWD},~\ref{GWS} and~\ref{CWs}, respectively, we briefly overview the current gravitational wave detectors and their sources, with particular attention devoted to a short discussion of the continuous wave signal.
Section~\ref{RecRes} is devoted to describe some of the main methodologies used in the searches for continuous gravitational waves, highlighting the most recent results obtained analyzing LIGO and Virgo data. 
A summary is finally reported in Sec.~\ref{Conclusion}.

\section{\label{GWD}Gravitational Wave Detectors}

The pioneer of gravitational wave detection was Joseph Weber in the early 1960s, who developed the first resonant mass detector and later also investigated laser interferometry~\cite{Weber}. From that date until today, the experiments that aim at the detection of gravitational radiation, planned in laboratories throughout the world, are in continuous progress~\cite{Auriga,Nautilus,Explorer,Allegro,Bonaldi,LeaciPRA,LeaciPRD,LeaciCQG,Virgo,Geo,TAMA,LIGOref1}. However, in this paper we consider only the efforts related to ground-based interferometric detectors.

The current world-wide network of gravitational wave detectors consists of the following Michelson-type kilometer-scale laser interferometers:
\begin{enumerate}
 \item{the French-Italian Virgo detector, at Cascina (Pisa, Italy), with 3 km arm length~\cite{Virgo}};
\item{the German-British GEO experiment, near Hannover (Germany), with an arm length of 600 m~\cite{Geo}};
\item{the Japanese TAMA, located in Tokyo (Japan), with 300 m arm length~\cite{TAMA}};
\item{the American LIGO project~\cite{LIGOref1}, that consists of three detectors working in unison; one at Livingston (Louisiana, USA), with an arm length of 4 km (LLO) and two in the same vacuum container at Hanford (Washington, USA), with an arm length of 4 km (LHO1) and 2 km (LHO2), respectively}.
\end{enumerate}

\begin{figure}[h!]
\begin{center}
\includegraphics[width=12cm, angle=0]{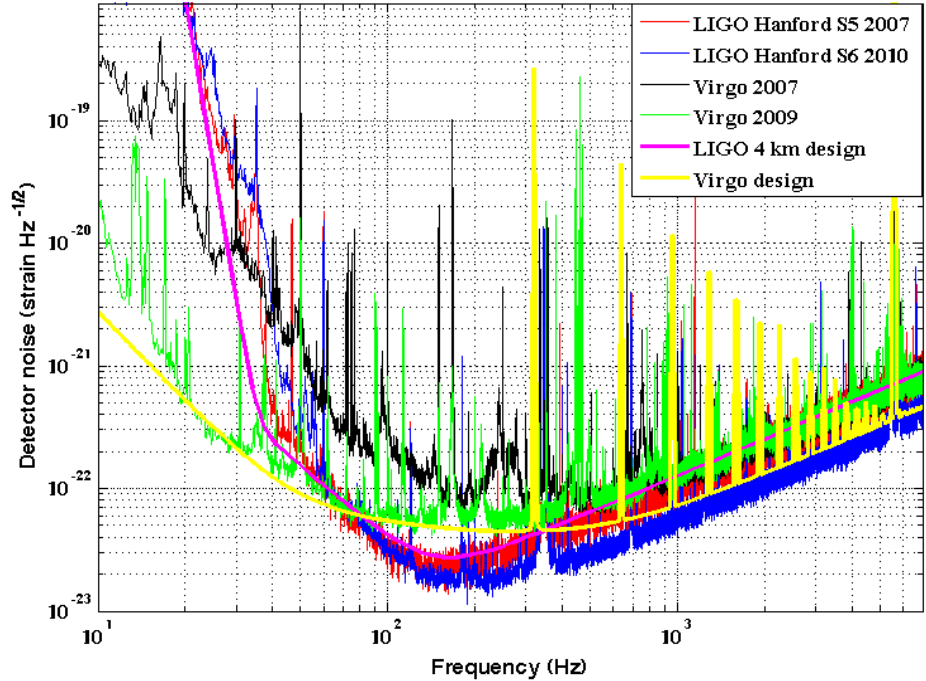} 
\caption{\label{Fig:AllSens} (color online). Strain sensitivity curves of the present gravitational wave interferometric detectors. In 2007 LIGO and Virgo reached their goal sensitivities in a wide frequency interval.}
\end{center}
\end{figure}

In 2007 LIGO and Virgo achieved their design sensitivities over a wide frequency range and today the performance of such interferometers is improved even more, as can be noticed in Fig.~\ref{Fig:AllSens}, where the noise spectra of such detectors is plotted versus the frequency. 
LIGO reached its design sensitivity with the fifth science run [S5 in short, started on 2005 November 4 (14) at LHO1 (LLO) and ended on 2007 October 1], but the sensitivity has continued to improve with time. In fact, with respect to S5, the sensitivity curve of a recently completed run, S6 (started on 2009 July 7 and ended on 2010 October 20), has a factor of 2 improvement above 300~Hz, as depicted in Fig.~\ref{Fig:AllSens}. Moreover, Virgo 2009 sensitivity measurements show a much better sensitivity than LIGO below $\sim40$~Hz (see Fig.~\ref{Fig:AllSens}).

The upgrade of LIGO and Virgo interferometers for their Advanced stage is currently underway, with the goal of improving the current strain sensitivity by a factor ten, with a thousandfold increase in the observable volume of space. In 2015 such interferometers will be operational and will gradually improve their sensitivity. At their target sensitivity, several gravitational wave events per year should be detected, opening thus the gravitational wave astronomy era.
 A future project for an interferometer of comparable sensitivity, LCGT (Large Scale Cryogenic Gravitational Wave Telescope), is going to be built in Japan~\cite{Lgct}.

Detectors of third generation, such as Einstein telescope, are currently in design phase~\cite{EinstTel}.

Moreover, space-borne detectors, such as LISA (Laser Interferometer Space Antenna)~\cite{LISA} and DECIGO/BBO (DECi-hertz Interferometer Gravitational wave Observatory and Big Bang Observer)~\cite{Decigo}, are designed to probe the 0.03~mHz to 0.1~Hz regime, bringing thus relevant astrophysical information at low frequencies and complementing the ground-based detectors.

In order to increase the baseline, it would be quite convenient to have a detector far away and out of the plane of other detectors in the USA and Europe. This would have a tremendous scientific impact and several advantages, such as to improve the ability to identify exactly where gravitational wave signals come from. The LIGO team is currently investigating some possibilities towards this direction.

\section{\label{GWS}Gravitational Wave Sources}

Different types of gravitational wave sources are expected to be observed by ground-based detectors. It is well-known that coalescence of compact objects constitutes an interesting source of high-frequency gravitational waves~\cite{Thorne1987}. In particular, the coalescence of NS-NS (NS, neutron star), NS-BH, BH-BH binary systems are expected to emit gravitational radiation in the kHz range~\cite{Thorne1995}. 
Such kind of sources are referred to as \textit{burst sources}, such as supernovae explosions, whose signals last for a very short amount of time, between a few milli-seconds and a few minutes. 

A \textit{stochastic background} of gravitational waves, either of cosmological or astrophysical origin, is also envisaged to exist. This consists of a random accumulation of signals from thousands or millions of individual sources.
Last, but not least, another class of gravitational wave sources is represented by rapidly rotating non-axisymmetric NSs, that are predicted to emit continuously a weak sinusoidal signal. In the next sections we limit ourselves just to the treatment of such \textit{continuous wave signals}.

The main mechanisms by which a NS can radiate gravitational waves consist of non-axisymmetric distortions in the solid part of the star (i.e. the case treated here, where the signal frequency $f$ is twice the star rotation frequency $f_{r}$)~\cite{Cutler2002,Ushomirsky}, free precession of the NS ($f=f_{r}$)~\cite{Stairs} and fluid \textit{r}-modes ($f \approx 4 f_{r}/3$)~\cite{Owen98,Andersson,NonAxNS1}. 

\section{\label{CWs}The continuous wave signal}

As already mentioned, continuous gravitational waves are expected to be produced by rapidly rotating NSs with non-axisymmetric deformations~\cite{NonAxNS1,NonAxNS5}.
The general form of a continuous gravitational wave signal is described by the following tensor metric perturbation:
\begin{equation}
\mathrm{\bf{h}}(t) = h_{+}(t) \mathrm{\bf{e}}_{+}(t) + h_{\times}(t) \mathrm{\bf{e}}_{\times}(t),
\end{equation}
where $h_{+}(t)$ and $h_{\times}(t)$ are the waveforms of the two orthogonal transverse polarizations, ``+'' and ``$\times$'', respectively and are given by
\begin{equation}
 h_{+}(t) = h_{0} \left(\frac{1 + \cos^{2} \iota}{2}\right) \cos \Phi(t), \qquad h_{\times}(t) = h_{0} \cos \iota \sin \Phi(t),
\end{equation}
with $\mathrm{\bf{e}}_{+,\times}$ representing the two basis polarization tensors~\cite{Misner}; $t$ is the time in the detector frame, $\iota$ is the inclination angle of the star's rotation axis with respect to the line of sight; $\Phi(t)$ is the signal phase function and $h_{0}$ is the amplitude expressed by
\begin{equation}
h_{0} = \frac{4 \pi^{2} G}{c^{4}} \frac{I_{zz} \epsilon f^{2}}{d}.
\end{equation}
The constant $G$ is the gravitational constant; $c$ represents the light speed; $I_{zz}$ is the star's principal moment of inertia (assumed to be aligned with its spin axis), $\epsilon$ is the equatorial ellipticity of the star~\cite{JKS}, $d$ is the distance to the star and $f$ represents the signal frequency. As the time-varying components of the mass quadrupole moment tensor are periodic with period half the star rotation period, the gravitational wave frequency $f$ is twice the rotation frequency $f_{r}$.

The detector response to a metric perturbation is given by the known relation
\begin{equation}
h(t) = F_{+}(t,\alpha,\delta,\psi)h_{+}(t) + F_{\times}(t,\alpha,\delta,\psi)h_{\times}(t),
\end{equation}
where $\alpha$ and $\delta$ are the source right ascension and declination, respectively, $\psi$ is the polarization angle of the wave and $F_{+,\times}$ are the detector antenna pattern functions for the two orthogonal polarizations~\cite{JKS}.

Assuming that all of the frequency's derivative, also denoted with the term of spin-down, is due to emission of gravitational radiation, we can relate $\epsilon$ to $f_{r}$ and $\dot f_{r}$~\cite{VelaApJ}:
\begin{equation}
h_{0}^{\mathrm{sd}} = 8.06 \times 10 ^{-19} I_{\mathrm{38}} \, d_{\mathrm{kpc}}^{-1} \sqrt{\frac{|(\dot f_{r}/\mathrm{Hz \, s^{-1}})|}{(f_{r}/\mathrm{Hz})}},
\end{equation}
where $I_{\mathrm{38}}$ is the star's moment of inertia in units of the canonical value $10^{38}$~kg~m$^{2}$ and $d_{\mathrm{kpc}}$ is the star's distance from the Sun in kiloparsec (kpc). This is referred to as \textit{spin-down limit} on the signal amplitude and represents an absolute upper limit to the amplitude of the gravitational wave signal that could be emitted by the star, where electromagnetic radiation is neglected. The spin-down limit on strain corresponds to an upper limit on the star's ellipticity, given by~\cite{VelaApJ}
\begin{equation}
\epsilon^{\mathrm{sd}} = 0.237 \left(\frac{h_{0}^{\mathrm{sd}}}{10^{-24}}\right) \, I_{\mathrm{38}}^{-1} \, (f_{r}/\mathrm{Hz})^{-2} d_{\mathrm{kpc}}.
\end{equation}

\section{\label{RecRes} Data analysis methods and recent results}

The way to search for continuous wave signals depends on how much about the source is known. There are different types of searches, briefly described in the text below and whose recent major results are also reported: 
\begin{enumerate}
\item{\textit{targeted searches}, where the source parameters (sky location, frequency, frequency derivatives) are assumed to be known with great accuracy (e.g. the Crab and Vela pulsars)};
\item{\textit{directed searches}, where sky location is known while frequency and frequency derivatives are unknown (e.g. Cassiopeia A, SN1987A, Sco X-1, galactic center, globular clusters)};
\item{\textit{all-sky searches} for unknown pulsars}.
\end{enumerate}

\subsection{\label{TargS} Targeted searches}

This kind of searches is computationally cheap and a fully coherent analysis, based on matched filtering over long observation time, is quite feasible~\cite{JKS}. 

The minimum signal amplitude that can be detected over a given observation time $T_{\mathrm{obs}}$, assuming a certain false alarm probability (typically of 1~\%) and a false dismissal probability (in general of 10~\%) is given by 
\begin{equation}\label{Eq:hominTS}
h_{0}^{\mathrm{min}} \approx 11 \sqrt{\frac{S_{h}(f)}{T_{\mathrm{obs}}}}.
\end{equation}
Equation~(\ref{Eq:hominTS}) is obtained by averaging over source and detector parameters and represents the sensitivity of a typical coherent search, with $S_{h}(f)$ being the detector noise power spectral density. Note that the precise value of the coefficient on the r.h.s of Eq.~(\ref{Eq:hominTS}) depends on the analysis method employed (see for example Ref.~\cite{Houghpaper} and references therein).

A search for continuous wave radiation from the Vela pulsar has been quite recently performed using data from the Virgo detector second science run (started on 2009 July 7 and ended on 2010 January 8)~\cite{VelaApJ}. The resulting upper limits on continuous gravitational wave emission have been obtained using methods that assume the gravitational wave emission to follow the radio timing. Assuming known orientation of the star's spin axis and value of the wave polarization angle, frequentist upper limits of $1.9 \times 10^{-24}$ and $2.2 \times 10^{-24}$, respectively, have been placed on the gravitational wave amplitude with 95~\% confidence level. An independent method, under the same hypothesis, produces a Bayesian upper limit of $2.1 \times 10^{-24}$ with 95~\%  degree of belief. These upper limits are well below the indirect spin-down limit of $3.3 \times 10^{-24}$ for the Vela pulsar, defined by the energy loss rate inferred from observed decrease in Vela's spin frequency, and correspond to a limit on the star ellipticity of the order of $10^{-3}$. Even assuming the star's spin axis inclination and the wave polarization angles unknown, the consequent results exhibit upper limits quite below the spin-down limit~\cite{VelaApJ}.

These recent results make Vela only the second pulsar for which the spin-down limit on gravitational wave emission has been beaten. The first pulsar for which this important result has been reached is the Crab pulsar~\cite{Abbott2008,Crab}. 

Another search worthy to be mentioned is the directed search for continuous wave signals from the non-pulsating NS in the supernova remnant Cassiopeia A over LIGO data. This search has established an upper limit on the signal amplitude over a wide range of frequencies which is below the indirect limit derived from energy conservation~\cite{CasA}. 

\subsection{\label{AllSkS} All-sky searches}

It is well-known that all-sky searches for gravitational waves from unknown pulsars over wide-parameter spaces are computationally limited. The reason is that one needs to search for unknown sources located everywhere in the sky, with signal frequency as high as a few kHz and with values of spin-down as large as possible. Long integration times, typically of the order of a few months or years, are needed to build up sufficient signal power.

The data analysis strategy used to extract the faint continuous wave signals from the interferometric noise data was derived in Ref.~\cite{JKS} and is given by the standard coherent matched filtering method, that is based on the \textit{maximum likelihood detection}. The resulting optimal coherent search statistic is the so-called $\mathcal{F}$-statistic.

 Fully coherent methods based on matched filtering are the approach used in analyses for continuous wave searches over wide parameter space. However, they become computationally undoable when very long data stretches (of the order of months or years) are used and a wide fraction of the parameter space is searched over, because of the increasing number of templates ~\cite{Jaran}. Therefore, different incoherent hierarchical methods have been proposed~\cite{BradyCrei,Houghpaper,CutlGh}.
In the hierarchical strategies, the entire data set is split into different shorter Fourier transformed data segments, which are then properly combined to account for Doppler shifts and spin-down. In other words, at first, every data chunk is analyzed coherently via matched filtering and afterward the information from the different segments is combined incoherently (that means that the phase information is lost). Three different methods have been developed that combine the results from the different segments incoherently, forming sums over power (``stack-slide''~\cite{BradyCrei,CutlGh} and ``PowerFlux''~\cite{AbbottPRD08powerf,AbbottPRL09powerflux} schemes) or weighted binary counts (``Hough transform''~\cite{Houghpaper,AbbottPRL05Hough,BadriHough}). The sums are then weighted according to the detector noise and antenna-pattern to maximize the signal-to-noise ratio.

The hierarchical methods are computationally faster than the standard coherent methods and have a comparable sensitivity.

In general, the whole data set, of duration $T_{\mathrm{obs}}$, is partitioned into $N$ smaller segments of duration $T_{\mathrm{coh}}$ each.

Given such $N$ data segments, the typical sensitivity of a continuous wave all-sky search is given by
\begin{equation}\label{Eq:hominAS}
h_{0}^{\mathrm{m}} \approx \frac{25}{N^{1/4}}\sqrt{\frac{S_{h}(f)}{T_{\mathrm{coh}}}},
\end{equation}
where the exact numerical factor depends on the specific hierarchical method employed (see for instance Ref.~\cite{Abbott2005} and references therein).

The output of a standard continuous wave hierarchical analysis is given by a set of candidates, i.e. points in the source parameter space with high values of a given statistic and which need a deeper study. Typically coincidences are done among the candidates obtained by the analysis over different data segments in order to reduce the false alarm probability~\cite{EatHearlyS5}. The surviving candidates can be then analyzed coherently over longer time baselines in order to discard them or confirm detection. 

Early LIGO data from S5 have been analyzed using two different methods. No gravitational waves could be claimed, but interesting upper limits have been placed. A first search used the first eight months of S5~\cite{Powerf09}, covering the full sky, the frequency band (50--1100)~Hz and a range of spin-down values between $-5 \times 10^{-9}$~Hz/s and zero. At the highest frequency the search would have been sensitive to the gravitational radiation emitted by a NS placed at 500~pc with equatorial ellipticity larger than $10^{-6}$.
Another search was performed over the first two months of S5 using the Einstein@Home infrastructure~\cite{EatHpro,EatHearlyS5}. The analysis consisted of matched filtering over 30~hours-long data segments followed by incoherent combination of results via a concidence strategy. The analyzed parameter space consisted of the whole sky, the frequency interval (50 --1500)~Hz and spin-down range between $-2 \times 10^{-9}$~Hz/s and zero. This search would have been sensitive to 90~\% of signals in the frequency band (125--225)~Hz with amplitude greater than $3 \times 10^{-24}$. The search sensitivity was estimated through Monte Carlo methods (injection of software simulated signals).

Three new Einstein@Home searches, covering the full two-years period of the LIGO S5 run, will be shortly published~\cite{LeaciOBS}.

\subsection{\label{EatH}The Einstein@Home project}

As already said, month-long coherent integration is necessary to accumulate a signal-to-noise ratio sufficient for detection. However, a powerful and effective method that allows us to use the longest possible coherent integration time, and thus improve the search sensitivity, is represented by distributing the computation through the volunteer computing project Einstein@Home~\cite{EatHpro}.
Such a project is built upon the BOINC (Berkeley Open Infrastructure for Network Computing) architecture~\cite{boinc}, namely a system that exploits the idle time on volunteer computers to solve scientific problems that require large amounts of computer power, such as to process data from gravitational wave detectors. At present, it provides roughly 300 TFlops of distributed computing resources.

A current Einstein@Home hierarchical search, analyzing data from S6 and using a new technique based on the $\mathcal{F}$-statistic global correlations~\cite{GlobCorr}, is expected to bring a relevant increase in terms of sensitivity. This is due to the longer coherent time baseline ($T_{\mathrm{coh}}$ = 60~hours) used in such a search.

\section{\label{Conclusion}Concluding remarks}
Despite huge efforts on several fronts (i.e. the improvement of the detector sensitivities and the employment of efficient gravitational wave data analysis algorithms), to date no direct gravitational wave detection has been made, but relevant upper limits on gravitational wave signal strength have been derived. Moreover, with the advent of advanced LIGO and Virgo detectors~\cite{AdvLIGO,AdvVirgo}, an amazing improvement in strain sensitivity, of a factor ten with respect to their initial configuration, is expected to be reached in a few years after 2015. At this point, the era of gravitational wave astronomy will definitely begin and the possibility of a first direct gravitational wave detection will become much more concrete. The employment of robust and hard-hitting gravitational wave data analysis techniques will be crucial at that time~\cite{PTDC,LeaciMethods,GlobCorr}.

\ack

The authors gratefully acknowledge the support of the United States
National Science Foundation for the construction and operation of the
LIGO Laboratory, the Science and Technology Facilities Council of the
United Kingdom, the Max-Planck-Society, and the State of
Niedersachsen/Germany for support of the construction and operation of
the GEO600 detector, and the Italian Istituto Nazionale di Fisica
Nucleare and the French Centre National de la Recherche Scientifique
for the construction and operation of the Virgo detector. The authors
also gratefully acknowledge the support of the research by these
agencies and by the Australian Research Council, 
the International Science Linkages program of the Commonwealth of Australia,
the Council of Scientific and Industrial Research of India, 
the Istituto Nazionale di Fisica Nucleare of Italy, 
the Spanish Ministerio de Educaci\'on y Ciencia, 
the Conselleria d'Economia Hisenda i Innovaci\'o of the
Govern de les Illes Balears, the Foundation for Fundamental Research
on Matter supported by the Netherlands Organisation for Scientific Research, 
the Polish Ministry of Science and Higher Education, the FOCUS
Programme of Foundation for Polish Science,
the Royal Society, the Scottish Funding Council, the
Scottish Universities Physics Alliance, The National Aeronautics and
Space Administration, the Carnegie Trust, the Leverhulme Trust, the
David and Lucile Packard Foundation, the Research Corporation, and
the Alfred P. Sloan Foundation.

\section*{References}

\end{document}